\documentclass[preprint,12pt,3p]{elsarticle}
\makeatletter
\def\ps@pprintTitle{%
 \let\@oddhead\@empty
 \let\@evenhead\@empty
 \def\@oddfoot{}%
 \let\@evenfoot\@oddfoot}
\makeatother

\usepackage{amsmath}
\usepackage{amssymb}
\usepackage[]{hyperref}

\begin{document}

\begin{frontmatter}

\title{Thermodynamics and Legendre Duality in Optimal Networks} 

\address[label1]{Department of Civil and Environmental Engineering, Princeton University, Princeton, New Jersey, 08540, USA}
\address[label2]{High Meadows Environmental Institute, Princeton University, Princeton, New Jersey, 08544, USA}
\address[label3]{Department of Biological and Agricultural Engineering, Texas A\&M University, College Station, TX, 77843, USA}
\address[label4]{Department of Environmental, Land, and Infrastructure Engineering, Politecnico di Torino, Corso Duca degli Abruzzi 24, 10129 Turin, Italy}
\address[label5]{Department of Mathematical Sciences, Politecnico di Torino, Corso Duca degli Abruzzi 24, 10129 Turin, Italy}
\address[label6]{INFN, Sezione di Torino, Turin 10125, Italy}

\cortext[cor1]{I am corresponding author}
\author[label1,label2]{Amilcare Porporato\corref{cor1}}
\ead{aporpora@princeton.edu}
\author[label3]{Shashank Kumar Anand}
\ead{skanand@tamu.edu}
\author[label3]{Salvatore Calabrese}
\ead{Salvatore.Calabrese@ag.tamu.edu}
\author[label4]{Luca Ridolfi}
\ead{luca.ridolfi@polito.it}
\author[label5,label6]{Lamberto Rondoni}
\ead{lamberto.rondoni@polito.it}

\begin{abstract}
Optimality principles in nonequilibrium transport networks are linked to a thermodynamic formalism based on generalized transport potentials endowed with Legendre duality and related contact structure. This allows quantifying the distance from non-equilibrium operating points, analogously to thermodynamic availability as well as to shed light on optimality principles in relation to different imposed constraints. Extremizations of generalized dissipation and entropy production appear as special cases that require power-law resistances and --for entropy production-- also isothermal conditions. Changes in stability of multiple operating points are interpreted as phase transitions based on non-equilibrium equations of state, while cost-based optimization of transport properties reveals connections to the generalized dissipation in the case of power law costs and linear resistance law, but now with typically unstable operating points which give rise to branched optimal transport.
\end{abstract}

\begin{keyword}
optimality \sep thermodynamics \sep networks \sep Legendre transform \sep dissipation \sep transport \sep entropy production
\end{keyword}

\end{frontmatter}

Regardless of whether variational approaches are epistemologically more objective, their underlying sense of design and purpose \cite{braithwaite1953scientific, chung2022thermodynamics} has permeated related research on thermodynamics and transport (see e.g., \cite{sewell1987maximum, rodriguez1997fractal, santambrogio2015optimal}). Optimization principles for macroscopic systems out of equilibrium have a long tradition dating back to Kelvin, Maxwell (minimum heat theorem for linear electrical circuits), Helmholtz, Dirichlet and Rayleigh (principle of least dissipation), followed by Onsager, Prigogine, Biot, Gyarmati, Ziegler, Paltrige, among many others \cite{van1999variational, niven2009steady, chung2022thermodynamics}. 

Confusion still remains, however, in part due to the existence of dual principles, which originate from a min-max structure and the Legendre-transform duality \cite{sewell1987maximum}. This echoes the equilibrium thermodynamic formalism \cite{callen1998thermodynamics}, where entropy is maximized in isolated systems, while different free energies are minimized for open systems, depending on the constraints imposed by the environment. Moreover, the conflation of entropy-production and dissipation extremizations is only justified in isothermal systems, where entropy production is proportional to dissipation (e.g., see Eq. \eqref{eq:entropro} below). Additionally, for power-law systems, the generalized transport potentials (GTPs; see Sec. \ref{sec:3}) are proportional to dissipation (and thus also to entropy production in isothermal conditions), further contributing to the idea that entropy production and dissipation are the general governing quantities for extremization, even for nonlinear systems. Thus, it is not surprising that variational principles based on entropy production often have proven misleading, as its extremization is problematic even in obtaining Fourier's heat-conduction law, where it leads to a $T^{-2}$ dependence of the heat conductance, instead of a constant value (e.g., \cite{barbera1999principle, kondepudi2014modern, martyushev2006problem}). 

The pioneering works of Millar and Cherry \cite{millar1951cxvi,cherry1963non} in the context of networks and then those of Edelen, Verhas and Presnov in nonequilibrium thermodynamics \cite{verhas2014gyarmati} have extended optimization principles to transport problems with nonlinear constitutive laws. Such principles depend on GTPs (Sec. \ref{sec:3}), constructed around Legendre transforms, having the product of generalized currents and driving forces as generators. As we will see, the resulting Lagrangians allow us to unify and clarify several previous optimization principles, such as the maximum or the minimum entropy production (MEP) and dissipation principles, which naturally follow with suitable constraints. 

While several contributions refer to optimizing currents and/or related driving forces, an interesting and somewhat independent line of research has focused on optimizing network configurations by adjusting the resistances/conductances. Following the classical approach of Murray \cite{murray1926physiological}, which in addition to dissipation also accounts for costs related to maintaining the transport system, several subsequent developments \cite{banavar1999size, bejan2004constructal, tero2007mathematical, bohn2007structure, rinaldo2014evolution} have shown how the long-term link between the flow and the optimal conductance induces a competition between branches, which in some cases produces unstable transport laws and branch pruning toward tree-like configurations. This corresponds to the case of optimal-channel-networks (OCNs) \cite{rodriguez1997fractal, bandle2017dido} and the related continuous models of evolution of fractal landscapes \cite{hooshyar2020variational, anand2023self}, as well as the models of branched optimal transport \cite{santambrogio2015optimal, facca2018towards}. 

In this work, we deal with optimal transport problems on discrete networks in stationary conditions, originating from extremizing Lagrangians formed with GTPs. In Sec. \ref{sec:2}-\ref{sec:networks} we show how such optimization principles allow unified treatment of several disconnected of previous results, yielding transport properties linking generalized fluxes and forces for both linear and nonlinear networks. Such transport laws are characterized by a Legendre duality similar to that of equilibrium thermodynamics. The resulting contact structure \cite{mrugala1991contact}, in particular, allows us to provide a measure of the distance from non-equilibrium steady state (NESS) transport process and naturally introduce gradient flows toward the network operating point. These analogies are further pursued in Sec. \ref{sec:stability}, where nonlinearities lead to multiple solutions of the constitutive laws but generalized dissipation potentials allow discriminating their stability, with analogies to non-equilibrium phase transitions. Sec. \ref{sec:optcond} deals with optimization of transport properties (i.e., the resistance laws); such efforts, because of the thermodynamic parallel between transport characteristics and thermodynamic material properties based on the second derivatives of Gibbs free energy (see Sec. \ref{sec:4}), can be seen as a macroscopic nonequilibrium branch of modern material science. The maintenance cost function results in apparent transport laws typically characterized by unstable behavior, which induces a competition among branches and result in optimal tree structures, related to branched optimal transport and evolutionary game theory \cite{tero2007mathematical,nowak2006evolutionary}. Again, in case of power laws, the optimization is equivalent to a minimization of generalized dissipation coupled to constraints of conservation of currents.

\section{Steady State Transport Laws}
\label{sec:2}

We consider transport phenomena on a directed network with resistive elements as a prototypical case of dissipative systems in non-equilibrium. 
Therefore, currents can only be generated by boundary conditions and no circulations in loops due to inner drivings are present \cite{Schnack}. The elements that form the $N$ branches of the network can be resistors in an electrical circuit, pipes in a water distribution network, 1D thermally conducting elements, and so on. We primarily focus on steady-state conditions. For a specific branch (no subscripts for notational simplicity), the balance equation for the transported quantity is
\begin{equation}
I=O=J 
\label{eq:massbal}
\end{equation}
where $I$ is the input to the branch and $O$ is the output, while $J$ is the flux or current through the link. Depending on the quantity being transported (e.g., electrical current, mass flow rate, internal energy, etc.), the current has different dimensions. At the nodes there may be a sink or source term. Virtual branches with ideal (current or force) sources can be added to obtain an augmented conservative network, thus avoiding node inputs and outputs without a corresponding branch. 

The transport or constitutive law connects the generalized forces $X$ and the current $J$, defining the so-called operating point
\begin{equation}
X = R(J)J  \ \ \ \ \ {\rm or} \ \ \ \ \ \ J=C(X)X,
\label{eq:operpoint}   
\end{equation}
expressed either in terms of resistance $R(J)$ or conductance $C(X)=1/R(J(X))$. 
In the linear case, 
in which $R$ and $C$ do not depend on $J$ or $X$, 
these are the well-known Poiseuille (i.e., laminar flow), Ohm, Fourier and Ficks laws; nonlinear generalizations are also of great interest, as they describe for example turbulent flows, state-dependent resistances, etc. We will often refer to the special case of power-law resistance,
\begin{equation}
 R = \Re J^{\alpha-1} \ \ \ \ \ \ \ \ \ \ \ \ \ {\rm or}
\ \ \ \ \ \ \ \ \ \ \  C = \Re^{\frac{1}{\alpha}} X^{\frac{1-\alpha}{\alpha}},
\label{eq:powerlaw}
\end{equation}
where $\Re$ is a problem-specific parameter. Thus, for example, the linear case $\alpha=1$ corresponds to laminar flow in pipes, while $\alpha=2$ corresponds to fully developed, hydraulically rough turbulent flow. The resistance parameter $\Re$ often contains information about the link cross-sectional geometry; for laminar flow in circular pipes, $\Re \propto r^{-4}$, where $r$ is the pipe radius, while for electrical resistors, $\Re \propto r^{-2}$. 

Multiplying Eq. \eqref{eq:operpoint} by the current gives a generalized energy balance
\begin{equation}
X J = R (J) J^2=C(X) X^2,
\label{eq:enbal}
\end{equation}
where the term on the left is the generalized power input (i.e., not necessarily an energy-input rate), while the ones on the right are the generalized dissipation rate (i.e., not necessarily an energy-dissipation rate). For the power law case \eqref{eq:powerlaw}, dissipation is
\begin{equation}
X J = \Re J^{\alpha+1}=\Re^{-\frac{1}{\alpha}}X^{\frac{1+\alpha}{\alpha}}.
\label{eq:enbalpower}
\end{equation}

Three things are worth noticing regarding these transport laws: 
\begin{itemize}
    \item {\it Steady state}. In case of time variability, the previous laws are no longer valid because of the presence of other processes related to storage terms, which are not included here. Similarly, `virtual' variations around the steady state give rise to extra components which will appear in suitable Lagrangian functions, as we will see in the next Section. One can, of course, associate a suitable time-like dimension to the relaxation of such perturbations toward steady state, to write the balance laws in the form of evolution equations. On the other hand, if the variation of the macroscopic quantities of interest is very slow compared to the evolution of such perturbations, a quasi-steady state generalized to time-dependent situations can be done by simply introducing time $t$ in them as if it were a parameter as in quasi-static thermodynamic transformations.
 
    \item {\it Single energy-mode networks}. We limit our discussion to the typical case of single energy mode, meaning that we only have one transported quantity \cite{oster1971tellegen}. Accordingly, we do not consider here cross-coupling between fluxes and different forces, as instead it is done in non-equilibrium thermodynamics, where dissipation is the sum of different fluxes driven by different, and possibly coupled, forces, $J_i=\sum_k C_{ik}X_i$, resulting in the famous Onsager reciprocal relationships $C_{ik}=C_{ki}$ \cite{martyushev2006maximum, Schnack}. Moreover, we only deal with constitutive laws that depend on the gradients of the driving potentials, but not on their actual values. 
    
\item {\it Focus on generalized dissipation, not entropy production}. When Eq. \eqref{eq:massbal} refers directly to energy transport, as in case of heat transfer, then \eqref{eq:enbal} does not refer to power or dissipation rate, but relates to what we call
`generalized' forms thereof. This notwithstanding, Eq. \eqref{eq:massbal} remains the starting point for extremization conditions. Only when Eq. \eqref{eq:enbal} refers to energy rates, do such terms relate to the entropy production rate,
    \begin{equation}
    \sigma=\frac{1}{T} R(J)J^2,
    \label{eq:entropro}
    \end{equation}
    provided $T$ is the uniform temperature in the branch. In all other cases, extremization of entropy production becomes misleading (see also \cite{jaynes1980minimum, martyushev2006problem}), as it does not correspond to extremization of the Lagrangian based on the generalized dissipation rates, defined in the next section.
\end{itemize}

\section{Variational Approaches based on Generalized Transport Potentials} \label{sec:3}

A remarkably general criterion exists for the optimization of both linear and nonlinear networks, based on generalized transport potentials (GTPs). These potentials were independently proposed for electrical networks by Millar and Cherry  \cite{millar1951cxvi, cherry1963non} and in nonequilibrium thermodynamics by Edelen, Verhas and Presnov (see \cite{verhas2014gyarmati} and references therein), but they are neither generally known nor they have been connected to the classical thermodynamic formalism.  

Focusing for now on a single branch, we consider virtual changes in either the flux $J$ or the force $X$, while the other variable is kept fixed at the operating point. This means allowing for perturbations of their values which do not necessarily satisfy the basic transport law \eqref{eq:operpoint}, with the goal of showing that the extremization of suitable Lagrangians is achieved at the operating point. This is analogous to the work of Onsager and Machlup on the linear laws \cite{OnsagerMachlup}, in which the extremization is performed in the large system limit. Depending on which quantity is 
varied, one obtains different principles which are connected by Legendre duality. One thing that is immediately clear for a single branch is that dissipation or entropy production alone are not sufficient to generate a functional that has an extremum at the operating point. 

In the space of currents -- also called flux representation -- the optimization principle is based on the extremization of the Lagrangian \cite{verhas2014gyarmati} \begin{equation}
\mathcal{L}_{J}(J)=XJ-\Phi(J),   
\label{eq:LagJ}
\end{equation}
where the GTP (or content)  \cite{millar1951cxvi} is
\begin{equation}
\Phi(J)=\int_0^J{X(J')dJ'}.  
\label{eq:content}
\end{equation}
This case corresponds to the Rayleigh-Onsager principle \cite{verhas2014gyarmati} (note, however, that Rayleigh defined the Lagrangian with the opposite sign, hence obtaining a minimum generalized dissipation principle, e.g., \cite{rayleigh1873some, doi2011onsager}) and is sometimes also referred to as the force-driven case, because $X$ is fixed and imposed externally, while $J$ is left to be found by optimization. The system's stable operating point \eqref{eq:operpoint} is obtained for the values of the current that maximizes $\mathcal{L}_J$. As we will see in Sec. \ref{sec:stability}, when the extreme of $\mathcal{L}_J$ is a minimum, the operating point is unstable to small dynamic perturbations.

The complementary or dual variational principle in the space of forces (also called Biot principle or force representation \cite{verhas2014gyarmati}) extremizes  
\begin{equation}
\mathcal{L}_X=XJ-\Psi(X),
\label{eq:LagX}
\end{equation}
where the GTP \cite{millar1951cxvi},
\begin{equation}
   \Psi(X)=\int_0^X{J(X')dX'},
   \label{eq:cocontent}
\end{equation}
is also called co-content.  

At the extreme of \eqref{eq:LagJ} the operating condition is $\mathcal{L}_J=\Psi$, while at the max of $\mathcal{L}_X$ it is value is $\mathcal{L}_X=\Phi$.
Thus, at the operating point, the generalized power input $XJ$ is the generator of the Legendre transformation between $\Psi$ and $\Phi$,
\begin{equation}
    \Psi(X)=XJ-\Phi(J),
    \label{eq:legendre}
\end{equation}
which provides a gradient structure to NESS processes, 
cf.\ e.g.\ \cite{bataille1979structuring}.
It is interesting to point out that, historically,  this terminology is connected to the so-called energy and co-energy -- or complementary energy -- duality in elasticity. The first appearance of this duality dates back to Castigliano's theorem for linear elasticity in 1873. From a thermodynamic point of view, the complementary energy is nothing but the Helmholtz free energy and thus the Legendre transform of the internal elastic energy \cite{langhaar1953principle,sewell1980complementary}. 

\begin{figure}[!hbtp]
\centering
\includegraphics[width=\linewidth]{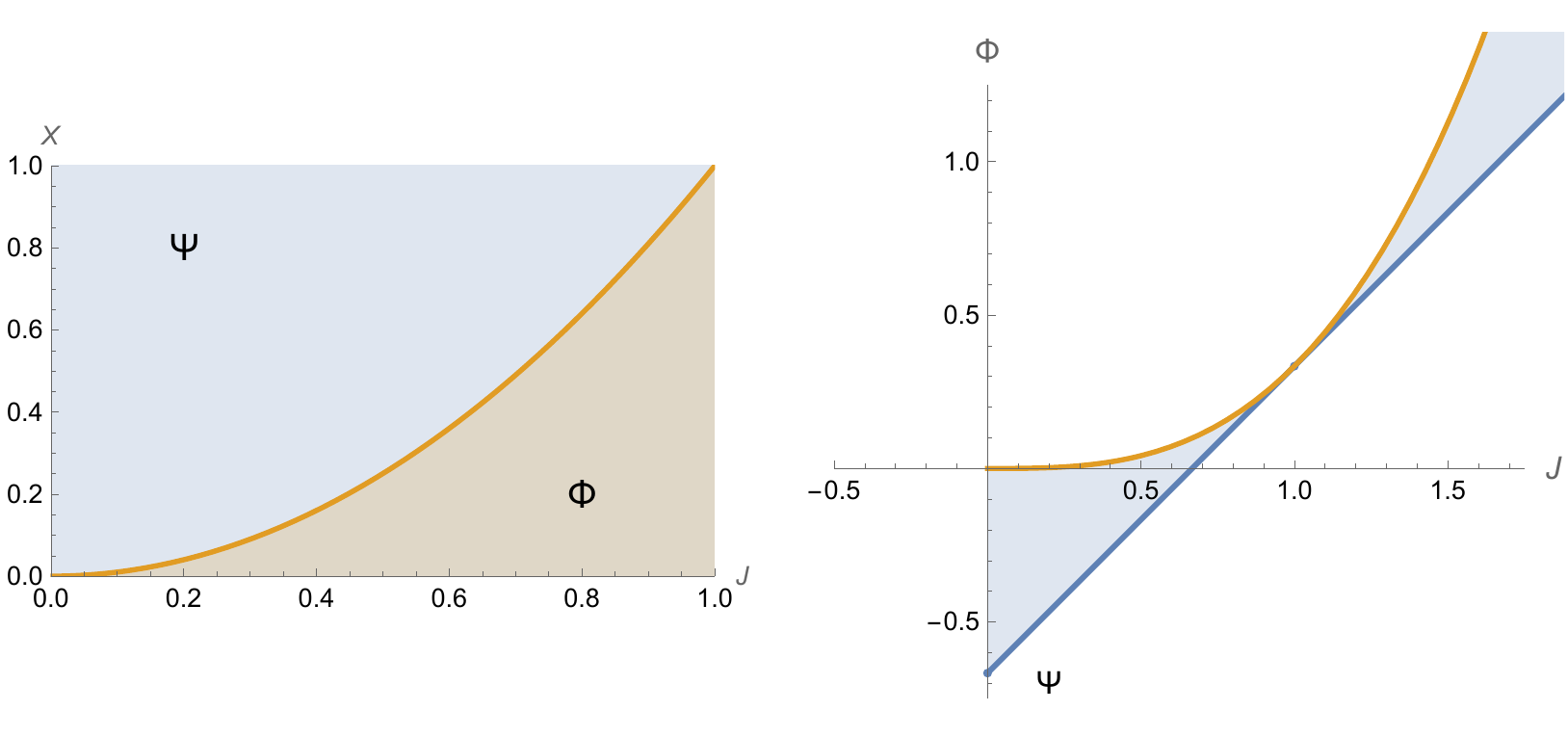}
\caption{GTPs (content and co-content) for nonlinear transport law, $\alpha=2$ and $\Re=1$ (left). On the right: related Legendre-transform structure with the shaded area representing the Lagrangian \eqref{eq:LagJ}.\label{fig:leg}} 
\end{figure}

In the linear case the GTPs, $\Phi$ and $\Psi$, are quadratic functions of currents and forces and, only for this case, they are equal numerically and then equal to half the generalized power input and the generalized dissipation rate. Moreover, for isothermal conditions, they are also proportional to the entropy production rate \eqref{eq:entropro}.  For the power-law case \eqref{eq:powerlaw} $\Phi$ and $\Psi$ are proportional to dissipation. A nonlinear power-law case ($\alpha=2$) is represented in Figure \ref{fig:leg}, which shows how the content and co-content differ for nonlinear constitutive laws (left) along with the Lagrangian $\mathcal{L}_J$ of Eq. \eqref{eq:LagJ}. For general nonlinear laws, the GPTs are no longer proportional to dissipation.

Gyarmati combined the two Lagrangians just presented, allowing for variations in both $X$ and $J$ \cite{gyarmati1970non,verhas2014gyarmati}. In this case, the operating point is given by the extremum of the Lagrangian 
\begin{equation}
\mathcal{L}=XJ-\Psi-\Phi.
\label{eq:Gyarmati}
\end{equation}
The operating point is always a maximum and equal to zero. This follows from the fact that, as we said before, $\Phi$ and $\Psi$ are Legendre transforms of each other at the functioning point; see Eq. \eqref{eq:legendre}. 

\begin{figure}[!hbtp]
\centering
\includegraphics[width=200pt]{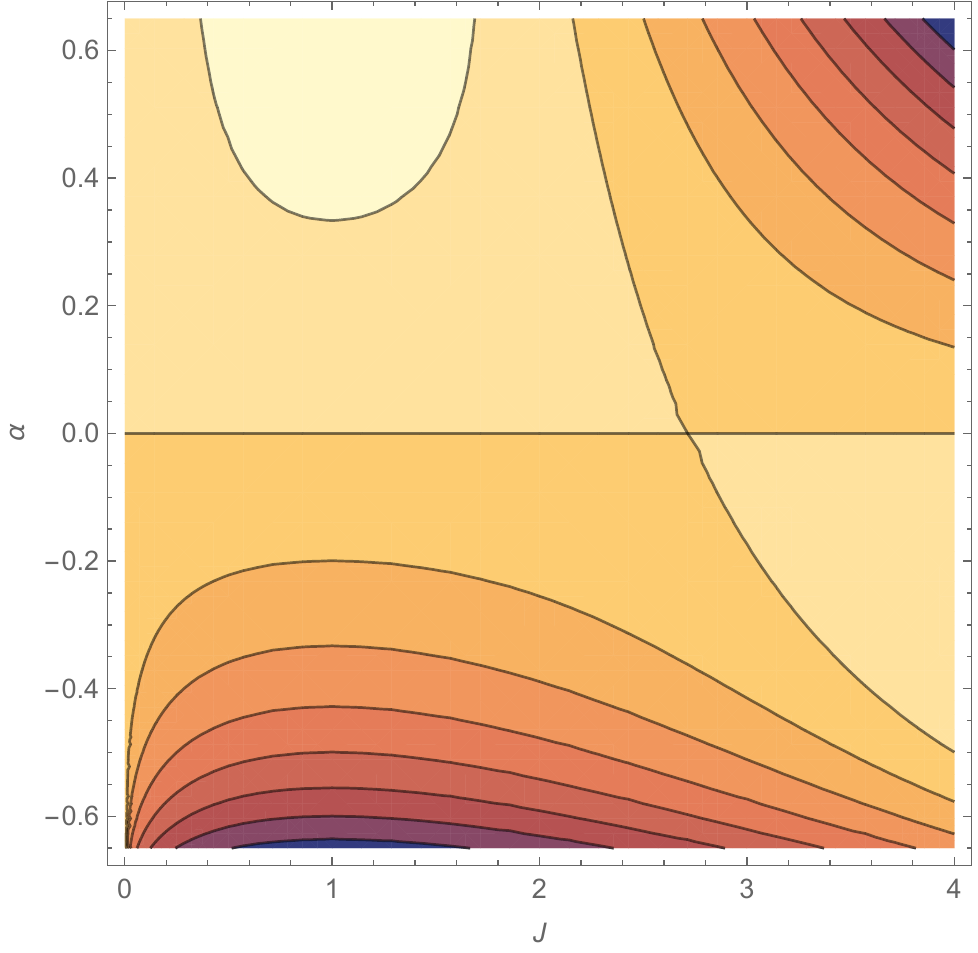}
\caption{Contour plot of the Lagrangian $\mathcal{L}_J$ \eqref{eq:LagJ} for power law resistance \eqref{eq:powerlaw} as a function of current $J$ and power-law exponent $\alpha$. Lighter colors mean higher values of the Lagrangian.\label{fig:saddle}}
\end{figure}

The constitutive law for the operating point, Eq. \eqref{eq:operpoint}, is typically a monotonic increasing function of $J$ and single valued. 
In case of more complex behaviors, Sewell \cite{sewell1980complementary} pointed out the link to Legendre transformation and catastrophe theory, while hysteresis was discussed in \cite{biolek2014co}.
For the case of multiple steady state values and instabilities see Sec. \ref{sec:stability}. Figure \ref{fig:saddle} shows the Lagrangian 
$\mathcal{L}_J$ \eqref{eq:LagJ} for the power-law resistance \eqref{eq:powerlaw} as a function of current and the power-law exponent \eqref{eq:powerlaw}. The extreme switches from a maximum for $\alpha>0$ to a minimum for $\alpha<0$ and shows the presence of a saddle point in the Lagrangian landscape. For $\alpha<0$ the operating point is unstable, leading either to zero or runaway currents for a given force $X$. We do not consider $\alpha\le -1$, because the GTPs do not converge. It is also not clear whether this condition actually corresponds to real transport processes.

\begin{figure}[!b]
\centering
\includegraphics[width=\linewidth]{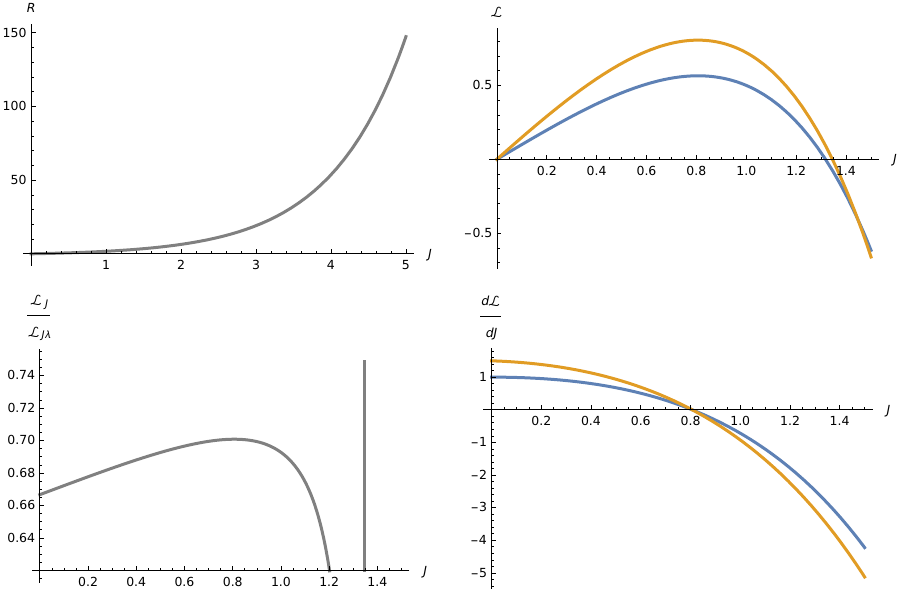}
\caption{Under exponential resistance law, $R(J)=e^J-1$; despite the difference between $\mathcal{L}_J$ (blue curve) and $\mathcal{L}_{J\lambda}$ (orange curve), their maximum is at the same value of the flux.\label{fig:lagrzie}}
\end{figure}

We have already mentioned that the principle obtained by maximizing the Lagrangian $\mathcal{L}_J$, Eq. \eqref{eq:LagJ}, can be seen as a generalization of the principles of Rayleigh and Onsager to nonlinear laws, although for generalized dissipation rates, rather than entropy rates. It is worth showing that it also corresponds to the principle of Ziegler \cite{ziegler2012introduction}, who obtained his principle by extremizing the generalized dissipation, constrained to satisfy the energy balance \eqref{eq:enbal} in steady state \cite{martyushev2006maximum}, that is
\begin{equation}
\mathcal{L}_{J\lambda}=R(J)J^2+\mu(XJ-R(J)J^2),   
\label{eq:LagZie0}
\end{equation}

or, with $\lambda=\mu-1$,
\begin{equation}
\mathcal{L}_{J\lambda}=XJ+\lambda(XJ-R(J)J^2),   
\label{eq:LagZie}
\end{equation}
where $\lambda$ is a Lagrange multiplier. The equivalence between \eqref{eq:LagJ} and \eqref{eq:LagZie} can be proven as follows. 
The condition $\partial \mathcal{L}_{J\lambda}/\partial \lambda=0$ 
for the Lagrangian \eqref{eq:LagZie} gives $XJ - R(J)J^2=0$, i.e.,
$X=R(J)J$. When the latter is substituted in $\partial \mathcal{L}_{J\lambda}/\partial J=0$, then, after little algebra, one obtains
\begin{equation}
    \frac{1}{\lambda}=1+\frac{J}{R}\frac{d R}{dJ}.
    \label{eq:lambda}
\end{equation}
Substituting \eqref{eq:lambda} into \eqref{eq:LagZie} to eliminate $\lambda$ yields a reduced Lagrangian which only depends on $J$. Differentiating it with respect on $J$ and setting it to zero leads to  
\begin{equation}
    \left( 2+\frac{R}{J}\frac{dJ}{dR}\right)(X-R(J)J)=0,
\end{equation}
which shows that the operating point \eqref{eq:operpoint} is always an extremum of the reduced Lagrangian, as it was for \eqref{eq:LagJ}. For power-law resistance \eqref{eq:powerlaw} one obtains $\lambda = \frac{1} {\alpha}$. In such a case, therefore, the Lagrangians $\mathcal{L}_J$ and the reduced $\mathcal{L}_{J\lambda}$ with the value of the Lagrange multiplier obtained from \eqref{eq:lambda} are proportional to each other. However, in the more general nonlinear case, they are not proportional, although they have the same extremum, as just shown. The case of an exponential resistance law is represented graphically in Figure \ref{fig:lagrzie}. It is not clear which of these two Lagrangians is more meaningful from a physical point of view, although the reduced Lagrangian obtained from \eqref{eq:LagZie} appears to have a more complicated mathematical form. From the point of view of the formalism is intended to describe, which is the constant NESS properties, both Lagrangians yield the same physics. Differences may arise if one extends the formalism to time-dependent properties. This is somewhat analogous to the ensemble equivalence in statistical physics, where microscopic fluctuations of macroscopic quantities differ according to the imposed constraints, while keeping the same expected values \cite{Gallavotti}.

Redoing similar calculations for the force-driven case, which is dual to the current-driven case \eqref{eq:LagZie}, the linear law $\alpha=1$ implies $\lambda=2$, a value which appears often in the literature (e.g., \cite{vzupanovic2004kirchhoff}; note however, that these authors emphasize maximizing entropy production, which only
in isothermal conditions is equivalent to maximizing dissipation, see Eq. \eqref{eq:entropro}.

\section{Thermodynamic Formalism for NESS}\label{sec:4}

From a thermodynamic point of view, the functioning point corresponds to a NESS described by the transport law \eqref{eq:operpoint}. In this Section we show that there is an interesting analogy between the formalism expounded in the previous section and the one of classical equilibrium thermodynamics \cite{callen1998thermodynamics}, which starts with the fundamental equation of Gibbs. For a one dimensional thermal system, the fundamental equation is $U=U(S)$, where $U$ is internal energy and $S$ is entropy. For the NESS transport problem this corresponds to $\Psi(X)$, which implies an analogy in the Legendre transforms, 
\begin{equation}
   F(T)=U(S(T))-TS
   \label{EquilLegend}
\end{equation}
and
\begin{equation}
    \Phi(J)=-\Psi(X(J))+JX.
\label{NoneqLegend}
    \end{equation}
The only difference is the sign convention used in thermodynamics. The equivalent of the entropy representation in thermodynamics here is a currents representation in $J$.
Thus the first derivatives define an analog for nonequilibrium of the equations of state:
\begin{equation}
\frac{\partial \Psi}{\partial X}=J
\label{eq:eqstate1}   
\end{equation}
\begin{equation}
\frac{\partial \Phi}{\partial J}=X,
\label{eq:eqstate2}   
\end{equation}
which correspond to the NESS transport (or constitutive) laws, see Eq. \eqref{eq:operpoint}.

Similarly, the second derivatives of the potentials, which in equilibrium thermodynamics define the material properties, here define the transport properties of the nonequilibrium system, i.e., the (nonlinear) conductances and resistances, respectively 
\begin{equation}
    \frac{\partial^2 \Psi}{\partial X^2}=C(X) \ \ \ \ {\rm and} \ \ \ \ \frac{\partial^2 \Phi}{\partial J^2}=R(J).
    \label{eq:traprop}
\end{equation}
These are new kinds of material properties, that arise in the description of non-equilibrium states. For the linear multivariate context, these are related by the famous Onsager reciprocal relationships \cite{martyushev2006maximum}), while in the thermodynamic analogy are Maxwell-type equations \cite{edelen1972nonlinear,bataille1979structuring}. 

The Gyarmati's Lagrangian \eqref{eq:Gyarmati} corresponds in equilibrium thermodynamics to the availability $A$ between two states of equilibrium. Specifically, the availability is the maximum work that can be extracted (by reversible transformations) from the system as it changes state; or equivalently, minus the minimum work needed to carry out reversibly the given change in thermodynamic quantities in the system (see, e.g., \cite{landau2013statistical, porporato2024thermodynamic}). For a thermal system (i.e., with constant mass and volume, as in the canonical ensemble), combining the first and second law of thermodynamics, the availability when going from a state $\hat{U}$, $\hat{T}$, $\hat{S}$ to a state at $U$, $T$, $S$, reads
\begin{equation}
A(T,\hat{S})=\hat{U}-U-T(\hat{S}-S)=F(T)-T\hat{S}-\hat{U}(\hat{S}).
\label{eq:availability1}
\end{equation}
Formally it is the difference between the Legendre transform and its definition when the arguments of the functions are considered independent; such a quantity is then zero only when $T$ corresponds to the slope of the fundamental equation at $S$, while in general for independent values of $T$ and $S$ it defines a non-zero surface and corresponds to a so-called contact structure \cite{mrugala1991contact}. The value of the availability, or similarly entropy production, by the Stodola-Guoy theorem, is zero at equilibrium, i.e.\ where $T\hat{S}-U$ is indeed the Legendre transform of $U(S)$, that is $F(T)$ because $T\hat{S}=TS$.
Thus, equivalently, treating both $X$ and $J$ as independent, moving from a flow configuration with $\hat{\Psi}$, $\hat{J}$, $\hat{X}$ to $\Psi$, $J$, $X$, one has 
\begin{equation}
-\mathcal{L}(J,\hat{X})=\Phi(J)-J \hat{X} +\hat{\Psi}(\hat{X}).
\label{eq:availability2}
\end{equation}

Thus, the quantity -$\mathcal{L}$ behaves like an availability for NESS, measuring the potential dissipation rate that would occur to equilibrate dynamically to the transport law, starting from different conditions of flux or force. Alternatively, it measures the extra dissipation of the transport process compared to the operating point, if the system is not in the optimal configuration or how far, energetically speaking, the system is from the constitutive law that one obtains at the maximum values of $\mathcal{L}$. To quantify this distance, following the thermodynamic analogy, the Lagrangian $\mathcal{L}$ could be employed to define a nonequilibrium metric, measuring the distance from 
NESS, including the distance of an equilibrium state from the NESS \cite{schlogl1985thermodynamic,mrugala1990statistical,brody1995geometrical,crooks2007measuring,niven2010jaynes}. This will be explored elsewhere.

\section{Reduced Lagrangians for Networks}\label{sec:networks}

The use of the Lagrangians \eqref{eq:LagJ} and \eqref{eq:LagX} can be extended to nonlinear networks simply by taking the sum over the network's branches. For the force-driven case, for example, summing over all the $N$ branches, the overall Lagrangian becomes
\begin{equation}
    \mathcal{L}_J=\sum_{i=1}^N {L}_{Ji}=\sum_{i=1}^N X_iJ_i-\sum_{i=1}^N \Phi(J_i).
    \label{eq:LagJnet}
\end{equation}
With this, the same variational properties and the thermodynamic analogy presented in the previous section both carry over to this multivariate case in the space of currents $J_i$. 

Often, however, this criterion is not very practical, because typically the driving forces are unknown in each branch. 
An alternative approach is offered by imposing the conditions given by the current continuity, i.e. the Kirchhoff's current laws (KCL). With this \cite{ercan2016kirchhoff}, the sum of the power input in each branch becomes a constant for given $X_i$, because the $M<N$ independent current remaining get multiplied by the Kirchhoff's voltage laws (KVL), i.e.,  
\begin{equation}
    \sum_{i=1}^N X_iJ_i=G,
    \label{eq:telle}
\end{equation}
where $G$, the generalized power absorbed from the external world, is assumed to be a known quantity. 

By substituting \eqref{eq:telle} in \eqref{eq:LagJnet}, the unknowns $X_i$ disappear, as desired; however, this also means that information on the network topology contained in the force vector at each branch is lost, and must be replaced by the continuity of currents at the nodes (i.e., Kirchoff's current law, KCL), thus also reducing the number of independent currents,
\begin{equation}
\mathcal{L}_J=G-\sum_{i=1}^N \Phi(J_i). 
\label{eq:LagJnetred1}\end{equation}  
When coupled to the KCL, a reduced Lagrangian is formally obtained as
\begin{equation}
\mathcal{L}^*_J(J_i) \ \ \ \ \ \ \  (i=1,...M),
\label{eq:LagJnetred2}\end{equation}  
where $M$ is the number of remaining independent currents. While this new reduced Lagrangian still has a maximum at the operating point, in general does not have the same contact structure associated to the Legendre transform. Of course, such a structure can be recovered by calculating the forces corresponding to the independent currents and then using the definition \eqref{eq:LagJ}, i.e., 
\begin{equation}
    \mathcal{L}^{**}_J=\sum_{i=1}^M X_iJ_i-\sum_{i=1}^M \Phi(J_i),
    \label{eq:LagJnetredrec}
\end{equation}
where now the $M$ forces are considered fixed. In the terminology of Callen \cite{greene1951formalism,callen1998thermodynamics}, one can see this as a microcanonical formulation with respect to all the non independent currents, which are not varied around the operating point, while the independent ones are allowed to vary as if it were in a canonical formulation with respect to those quantities \cite{porporato2024thermodynamic}.

\begin{figure}[!b]
%\centering
\includegraphics[width=400pt]{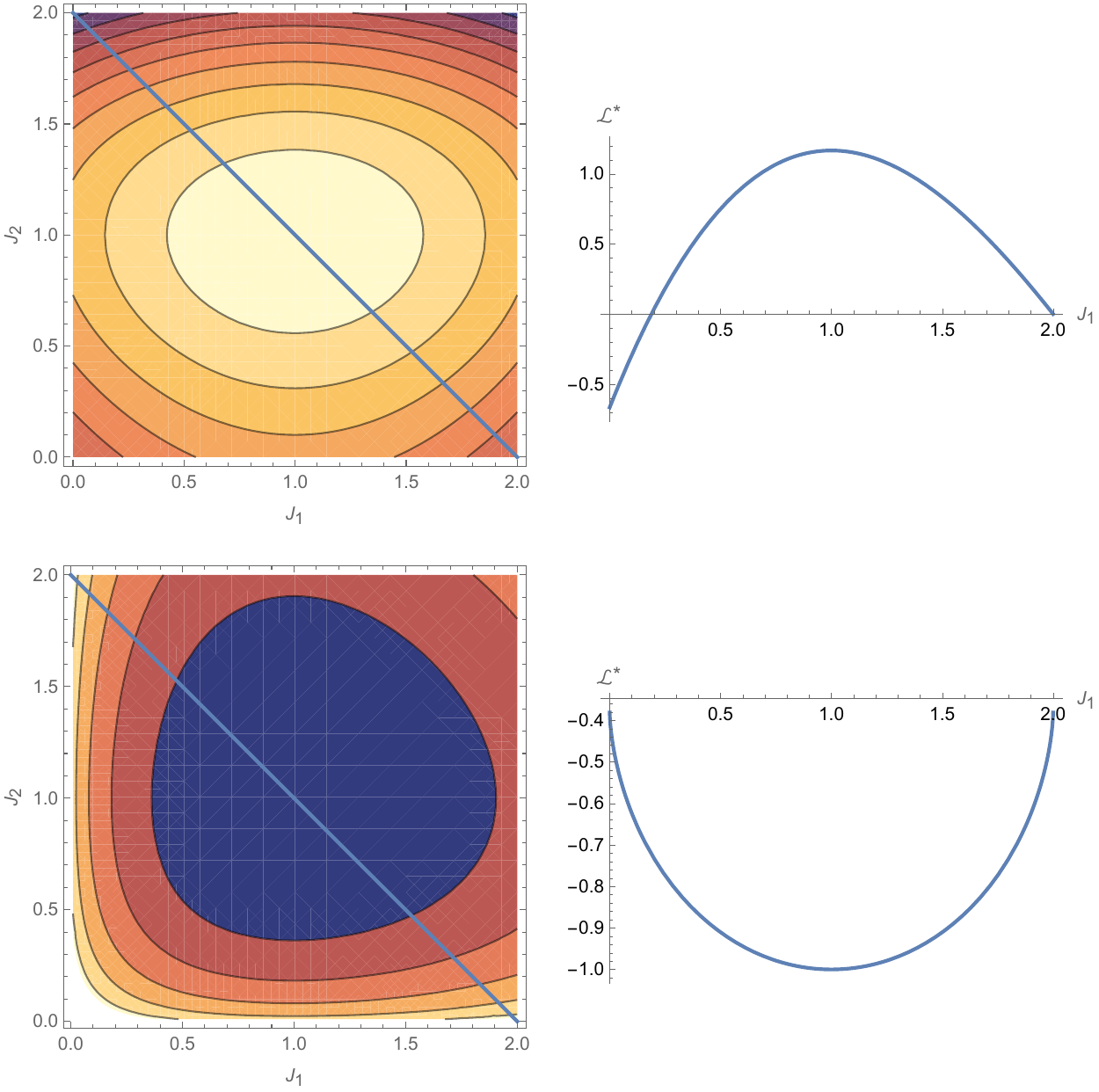}
\caption{Parallel pipes. Contour plots for the complete Lagrangians $\mathcal{L}_J$ (left panels) and reduced Lagrangians $\mathcal{L}^*_J$ (right panels). Lighter/darker colors denote higher/lower values of the Lagrangian. Top panels refer to the stable case with different power laws ($\alpha_1=1$ and $\alpha_2=2$); bottom panels refer to the unstable case ($\alpha=-1/3$ for both pipes).}
\label{fig:parpipes}
\end{figure}

Equations \eqref{eq:LagJnetred2} and \eqref{eq:LagJnetredrec} imply the presence of a potential for the network in the space of currents whereby the approach to the operating point is a result of a gradient flow as, e.g., 
\begin{equation}
\label{eq:gradlaw}
    \frac{\partial J_i}{\partial t}=\frac{\partial \mathcal{L}^{**}_J}{\partial J_i},
\end{equation}
for $i=1,...,M$, where $t$ is a time-like variable. The convergence to an absolute maximum or local minima depends on the convexity or concavity of the Lagrangian of the various branches (see e.g., Figure \ref{fig:saddle}).

The case of two parallel pipes provides a very simple illustration of the generality of this approach. We consider the case of different nonlinear (power-law) transport laws, which was also analyzed in an interesting paper by Niven \cite{niven2010simultaneous}, where he provided an analysis of extremum principles based on entropy production, pointing out the limitations and differences of various entropy-based criteria for this case. As shown graphically here in Figure \ref{fig:parpipes}, the optimization of the Lagrangians \eqref{eq:LagJnet} and \eqref{eq:LagJnetredrec} obtained using generalized dissipation potential provides a solution also in case of different power laws, going beyond the limitations of the other principles based on MEP and dissipation. The left figure shows the complete Lagrangian \eqref{eq:LagJnet}, compared with the reduced Lagrangian $\mathcal{L}^*$ of Eq. \eqref{eq:LagJnetred2}, resulting from the substitution $J_2=J-J_1$ and $X_1=X_2=X$, with $J=2$ and $X=1$. It is also instructive to analyze the unstable case with negative exponent, $\alpha=-2/3$. For this, the flows in the two pipes 'compete' for the total flow, as it happens in evolutionary games (e.g., \cite{nowak2006evolutionary}; see also \cite{tero2007mathematical}), giving rise to a single pipe configuration carrying the entire flow. A similar situation with multiple local extremes will present itself again in Sec. \ref{sec:optcond}, dealing with the selection of optimal conductances in the case of sub-additive cost functions producing optimal tree networks.

\section{Maximum or Minimum Entropy Production?}

For the specific case of power-law resistances, the previous Lagrangians acquire a special form related to the generalized dissipation. Here we show how, in this case, both the maximum entropy production and the minimum dissipation principles follow from the general Lagrangian, Eq. \eqref{eq:LagJnet}, using different constraints. For power laws \eqref{eq:powerlaw} the total content is proportional to the total virtual dissipation $\mathcal{D}$, i.e., 
\begin{equation}
  \sum_{i=1}^N \Phi(J_i)=\frac{1}{\alpha+1}\sum_{i=1}^N \Re_i J_i^{\alpha+1}=\frac{\mathcal{D}}{\alpha+1}.
\end{equation}
Then, with the use of KCL, the reduced Lagrangian becomes
\begin{equation}
    \mathcal{L}^*_J=G-\frac{\mathcal{D^*}}{\alpha+1},
    \label{eq:redlagpower}
\end{equation}
where $\mathcal{D^*}$ is the reduced virtual dissipation, obtained by varying only the $M<N$ remaining independent currents. As a result, at the operating point, when such currents satisfy KVL, the reduced virtual dissipation is minimized, while the reduced Lagrangian above is maximized. In isothermal conditions, these principles directly translate into entropy-production extremizations: the maximization of $\mathcal{L}^*_J$ is Onsager's and Ziegler's principle of maximum entropy production \cite{martyushev2006maximum}, while the dissipation minimization corresponds to Prigogine's minimum entropy production principle \cite{kondepudi2014modern} (and therefore also the minimum heat theorem of Maxwell and the least energy dissipation of Rayleigh). 

It should be clear at this point that similar results can also be obtained starting with the formulation \eqref{eq:LagZie} with the Lagrange multiplier \eqref{eq:lambda}. In fact, for the power-law case, adding the KCL as a constraint,  ($\lambda=1/\alpha$), one obtains
\begin{equation}
\mathcal{L}_{J\lambda}^*=G+\frac{1}{\alpha}G-\frac{1}{\alpha}\mathcal{D^*}=\frac{\alpha+1}{\alpha}\left( G-\frac{\mathcal{D^*}}{\alpha+1}\right),
\end{equation}
which has the same variational properties as \eqref{eq:redlagpower}, being proportional to it. It is important to stress that for non-power-law resistances, the GTPs are no longer proportional to generalized dissipation; as a result, dissipation or entropy production principles no longer hold.

\section{Stability and Dynamic Phase Transitions}\label{sec:stability}

The transition between different dynamic regimes as the driving force increases has often been interpreted in an {\it ad hoc} manner using optimality theories, based on dissipation and entropy production \cite{harsha2019review, kawazura2012comparison, martyushev2007some, niven2010simultaneous}. These cases typically include a transition from a linear regime at low flows (or forces) to a nonlinear regime at strong flow (or forces), with examples ranging from turbulence transition in pipes \cite{martyushev2007some,niven2009steady}, to
formations of different crystals \cite{hill1990entropy,martyushev2007some}, onset of Rayleigh-Bernard convection \cite{ozawa2001thermodynamics} and the  transition from abiotic to biological systems \cite{vallino2010ecosystem, vallino2016thermodynamics}. Here we show that the stability of the constitutive law is predicted by the Lagrangians \eqref{eq:LagJ} and \eqref{eq:LagX}, introduced in Sec. \ref{sec:3}, depending on whether the extreme is a maximum (stable) or a minimum (unstable). These can be considered as dynamic phase transitions, in analogy with phase transitions determined by thermodynamic potentials like the Gibbs free energy \cite{callen1998thermodynamics, ge2009thermodynamic}. 

For a flux-driven 1D system, the unsteady evolution around the operating point may be described formally by an evolution equation of the type
\begin{equation}
\label{eq:lj_evo}
    \frac{dJ}{dt}=\frac{\partial \mathcal{L}_J(J)}{\partial J}=X-\frac{\partial \Phi(J)}{\partial J}=X-R(J)J,
\end{equation}
where information on the timescales of approach to the NESS is not contained in the generalized dissipation potentials. In case of multiple extreme points, the solution is determined by the type of extreme, as in the system of Kawazura, Yoshida and coworkers \cite{yoshida2008maximum, yoshida2014bifurcation, kawazura2010entropy, kawazura2012comparison}. 
This nonlinear system, which is not in power-law transport form, clearly shows that neither the dissipation nor the entropy production can be used as criteria for optimality, differently from what is maintained by other principles in the literature, such as the Glansdorf-Prigogine \cite{glansdorff1964general} 
or the MEP \cite{martyushev2006maximum}. Rather, the stability and selection of the NESS should be based on the Lagrangians \eqref{eq:LagJ} and \eqref{eq:LagX} and their second variations. As a result, these nonequilibrium transitions \cite{ge2009thermodynamic} are recast as a switch in stability between solutions of a `dynamic equation of state' (i.e., \eqref{eq:eqstate1} and \eqref{eq:eqstate2}, with multiple roots. When integrated, such an equation gives the corresponding potential, whose second variation at the extremes provides the stability of the solutions. Crucially, the integration done to construct the potentials, see Eqns. \eqref{eq:content} and 
\eqref{eq:cocontent}, implies that, in general, the stability criteria cannot be local (i.e., based only on the information on the neighborhood of the NESS or at a point in space for continuous systems), as shown by the integration over the entire domain of the current and flux in \eqref{eq:content} and \eqref{eq:cocontent}, respectively. This is in agreement with Landauer's objection \cite{landauer1975stability} to the locality of the Glansdorf-Prigogine criterion 
\cite{glansdorff1964general} (nonlocality also arises in statistical mechanics descriptions of NESS, even in the presence of local thermodynamic equilibrium \cite{Bodin, Derrida, Bertini}.)

In the next two subsections, we discuss the nonlinear model of Kawazura and coworkers \cite{kawazura2012comparison,yoshida2014bifurcation}, first for the case of constant forces and then for the case of constant current.

\subsection{Self-Organized Plasma: Force-Driven Case}

\begin{figure}[!hbtp]
\centering
\includegraphics[width=\linewidth]{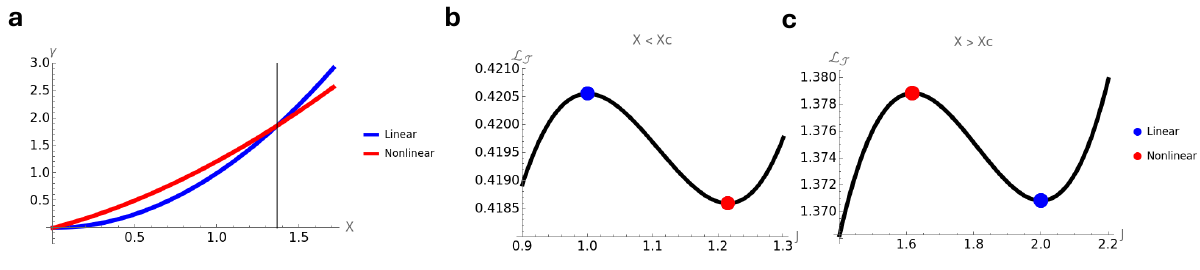}
\caption{\label{fig:lj_yoshida}Self-organized boundary layer in a force-driven plasma flow. (a) Dissipation for the linear (blue) and nonlinear (red) branches, where the vertical line marks the critical force value $Xc$. (b,c) $\mathcal{L}_J$ as a function of $J$, when $X<Xc$ (panel b) versus when $X>Xc$ (panel c). For these results, we used $R_0=1$, $a=3$, and $T_0=1$, which yields $X_c = 1.366$. Panel b is for $X=1 \left(<X_c \right)$ and panel c plots $\mathcal{L}_J$ for $X=2 \left(> X_c \right)$}
\end{figure}

We consider a nonlinear resistance induced by a self-organized boundary layer in a plasma flow \cite{kawazura2012comparison,yoshida2014bifurcation}. The problem is similar to the case of turbulence transition \cite{martyushev2007some,niven2009steady}, where the resistance law becomes nonlinear with faster increase as a function of flow as the regime changes from laminar to turbulent. It comprises a constant resistor $R_0$ in series with a nonlinear one, 
$$R_1=a J\left(\frac{T_0}{T_0+R_0J
}-\frac{T_0}{X+T_0}\right),$$
where $a$ is a parameter and $T_0$ is a reference temperature. The driving force is $X=T-T_0$. The non-equilibrium `equation of state' characterizing the steady state can be expressed in terms of an equivalent resistance,
\begin{equation}
\label{eq:x_yoshida}
X = \frac{\partial \Phi}{\partial J} = \left( R_0 +a\left(\frac{T_0}{T_0+R_0J
}-\frac{T_0}{X+T_0}\right)J\right)J.    
\end{equation}
Since this can be written as
\begin{equation}
    (X-R_0J)\left(\frac{aT_0J^2}{R_0(X+T_0)J+X+T_0}-1\right)=0,
\end{equation}
the two physically possible currents are easily found to be
\begin{equation}
    J=\left\{\begin{array}{l}
       X/R_0 \\[8pt]
\cfrac{R_0 (X+T_0) +
\sqrt{(R_0(X+T_0))^2+4aT_0^2(X+T_0)}}{2aT_0}, 
    \end{array}\right.
\end{equation} 
which cross at
\begin{equation}
    X_c = \frac{T_0 \sqrt{a T_0}}{\sqrt{a T_0}-R_0}-T_0.
\end{equation}
The actual stable solution is revealed by the potential $\mathcal{L}_J$, resulting from Eq.\ \eqref{eq:x_yoshida}. This is expressed by
\begin{eqnarray}
\mathcal{L}_J &=& \int^J \left(X - \frac{\partial \Phi}{\partial J'}\right)dJ'= XJ - \Phi(J) \\
&=&\frac{1}{6} J \left(\frac{2 a J^2 X_0}{X_0+X}-\frac{3 J \left(a X_0 + R_0^2\right)}{R_0} + 6 \left(\frac{a X_0^2}{R_0^2}+X\right)\right) - \frac{a X_0^3 \ln (J R_0 + X_0^2)}{R_0^3} \nonumber 
\end{eqnarray}
Setting the dissipation $\Upsilon=XJ$, the observed solution corresponds to the point in which $\mathcal{L}_J = \Upsilon-\Phi(J)$ has a maximum, so that the linear solution is selected for $X<X_c$ and the nonlinear branch is selected for the force greater than the critical value. Figure \ref{fig:lj_yoshida}(b,c) shows the plot of $\mathcal{L}_J$ for the system, where the stability of the solutions changes as $X$ crosses the critical value. Thus, from a dynamical system viewpoint, the evolution of the macroscopic system obeys an equation of the following form 
\begin{equation}
   \tau_X \frac{dX}{dt}= -\frac{\partial \mathcal{L}_J}{\partial J}.
   \label{eq:lj_evol}
\end{equation}
Interestingly, if one does not look at the potential $\mathcal{L}_J$, but only looks at the dissipation rate ($\Upsilon = XJ$) or the entropy production \cite{martyushev2007some}, then the selected solution appears to correspond to a minimum of the dissipation rate (Figure \ref{fig:lj_yoshida}(a)), which is the opposite of what takes place in the flux-driven case discussed next.

\subsection{Self-Organized Plasma: Flux-Driven Case}

\begin{figure}[!hbtp]
\centering
\includegraphics[width=\linewidth]{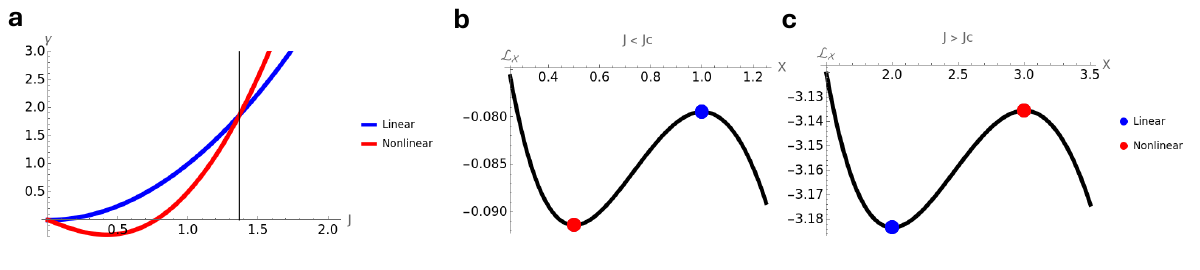}
\caption{\label{fig:lx_yoshida}Self-organized boundary layer in a flux-driven plasma flow. (a) Dissipation for the linear (blue) and nonlinear (red) branch, where the vertical line marks critical flux value $Jc$. (b) $\mathcal{L}_X$ as a function of $X$, when $J<Jc$ (panel b) versus when $J>Jc$ (panel c). For these results, we used $R_0=1$, $a=3$, and $T_0=1$, which yields $J_c = 1.366$. Panel b plots $\mathcal{L}_X$ for $J=1 \left(<J_c \right)$ and panel c displays $\mathcal{L}_X$ for $J=2 \left(> J_c \right)$.}
\end{figure}

For the flux-driven case, the two steady-state solutions are
\begin{equation}
    X = \left\{\begin{array}{l}
       R_0 J \\
       \frac{a J^2 T_0}{J R_0+T_0}-T_0
    \end{array}\right.
\end{equation} 
and cross at a certain driving flux,
\begin{equation}
    J_c = \frac{T_0}{\sqrt{a T_0}-R_0}.
\end{equation}
Plotting the dissipation rate, one can say that the solution is selected such that the dissipation rate is maximum (see Figure \ref{fig:lx_yoshida}(a)), which is the opposite of what is found in the force-driven case.

The unified criterion is provided again by the generalized dissipation potential. One can reformulate Eq. \eqref{eq:x_yoshida} as
\begin{equation}
J = \frac{\partial \Psi}{\partial X}= \frac{X}{R_0} - \frac{a}{R_0} \left(\frac{T_0}{T_0+R_0J}-\frac{T_0}{X+T_0}\right)J^2,
\end{equation}
from which one obtains the potential  
\begin{equation}
\mathcal{L}_X = \int^X \left(J-\frac{\partial \Psi}{\partial X'}\right)dX'= XJ - \Psi(X) = \frac{X \left(\frac{2 J \left(J \left(a T_0+R_0^2\right)+R_0 T_0\right)}{J R_0+T_0}-X\right)-2 a J^2 T_0 \ln (T_0+X)}{2 R_0}.
\end{equation}
As seen in Figure \ref{fig:lx_yoshida}, the operating point corresponds to the point where $\mathcal{L}_X$ has a maximum, just like the force-driven case for $\mathcal{L}_J$. Again, an evolution equation can be written down, as Eq. \eqref{eq:lj_evol},
\begin{equation}
\label{eq:lx_evol}
   \tau_X \frac{dX}{dt}=-\frac{\partial \mathcal{L}_X}{\partial X},
\end{equation}
where the stability of the solutions changes from linear to nonlinear branch with $J$ being smaller or greater than $J_c$, see Figure \ref{fig:lx_yoshida}(b,c). 

In summary, this nonlinear case clearly shows that dissipation and entropy production by themselves can not be used as discriminant for the stable solutions of the system, which is instead provided by the use of GTPs and the related Lagrangians, $\mathcal{L}_J$ and $\mathcal{L}_X$.
 
\section{Optimal Transport Properties\label{sec:optcond}} 

Another form of optimality may be achieved by allowing for variations in transport properties, i.e., the resistances in \eqref{eq:operpoint}. While, physically speaking, this form of optimization is very different from the one discussed before in this paper, it is interesting here to point out the formal similarity with the previous optimizations, especially in relation to the minimization of dissipation, which follows in case of power-law cost functions, and the common presence of unstable apparent transport laws that lead to tree-type networks. It is also worth mentioning that, in our analogy with thermodynamical systems, transport properties play the same role in the extended thermodynamics of NESS outlined in Sec. \ref{sec:4} as the material properties do in thermodynamics, where they are related to the curvatures of the fundamental equation and its Legendre transforms (e.g., Gibbs free energy), see e.g., \eqref{eq:traprop}. Thus, adjusting transport properties also has analogies with the efforts in modern material science, where properties are designed and manufactured to obtain specific goals \cite{torquato2010optimal}. The problem of optimal transport properties is also connected to the adaptation and evolution of natural systems (including biological, where the optimization is often framed as a maximization of a fitness measure \cite{waldherr2015dynamic, franklin2020organizing, calabrese2022soil}) and the design of engineered ones, as well as to optimal transport problems {\it a la} Monge-Kantorovich \cite{santambrogio2015optimal, facca2018towards, anand2023self}.

One of the simplest examples of this topic is the design optimization of a pipe cross-section for a given flow rate or driving potential, a classical integral variational problem (e.g., Dido's maximum area problem \cite{bandle2017dido, witelski2015methods}). Bejan used it in support of his constructal theory \cite{bejan2004constructal}, invoking the variation of shape parameters, e.g.\ the number of sides of a polygonal cross section with given area, to achieve the circular cross-section of optimal conductance. For this problem, the duality or min-max form of the optimization \cite{sewell1987maximum} is immediately apparent, for the circular cross-section has maximum flow and dissipation for given driving force, $X$, while it has minimum driving force, and therefore also dissipation, for given flow $J$. 

Focusing here on Murray's well-known optimization of biological transport \cite{murray1926physiological}, we add a maintenance cost to the energy dissipation by the flow. This produces optimal conditions even for a single pipe, without need to impose constrains on total cross-sectional area. The optimal link between resistance and current provides a criterion that can then be applied to obtain optimal configurations of transport between assigned inputs and outputs and intermediate nodes \cite{rodriguez1997fractal, bohn2007structure, anand2021minimalist, giaccone2025coexistence}, ranging from congested to branched transport, with clear parallels to optimal channel networks and optimal transport problems \cite{hooshyar2020variational,facca2018towards}.  For a given current in a branch, the cost function is equal to the sum of dissipation and a construction/maintenance cost, 
\begin{equation}
    \mathcal{M}_i=R_i(J_i, \Re_i)J_i^2+f(\Re_i),
    \label{eq:lagmu0}
\end{equation}
where $\Re$, as in (\ref{eq:powerlaw}), is again a parameter related to the branch property (e.g., the radius). Since the optimization of the transport properties is usually a slow process, the flow in the branch may be assumed to be already at the operating NESS point \eqref{eq:operpoint}, as discussed in Sec. \ref{sec:3}. This means that the optimization between $X_i$ and $J_i$ has already taken place and we are now optimizing the values of resistances by adjusting the cross-sectional properties of the branch. Because dissipation increases while the pipe costs typically decrease with resistance, the overall cost $\mathcal{M}_i$ has a minimum at an optimal resistance, which also corresponds to an optimum value of the force. 

For the special case of power-law resistance and cost,
\begin{equation}
\mathcal{M}_i=\Re_iJ_i^{\alpha +1}+\frac{k}{\Re_i^\beta},
\label{eq:lagmu}
\end{equation}
differentiating with respect to $\Re_i$, for constant current, yields an optimal resistance, one-to-one related to the current, i.e.,
\begin{equation}
J_i=(k\beta)^{\frac{1}{\alpha+1}}\Re_i^{-\frac{1+\beta}{1+\alpha}}.
\label{eq:optj}
\end{equation}
or
\begin{equation}
\Re_i=(k\beta)^{\frac{1}{\beta+1}}J_i^{-\frac{1+\alpha}{1+\beta}}.
\label{eq:optr}
\end{equation}
Thus, upon optimizing the resistance parameter, the overall relationship between current and forces can be represented by an 'apparent' resistance law with an exponent
\begin{equation}
\check{\alpha}=\frac{\beta-\alpha}{1+\beta};
\end{equation}
for Murray ($\alpha=1$ and $\beta=1/2$) $\check{\alpha}=-1/3$, which corresponds to an unstable case (e.g., see the case with $\alpha=-1/3$ in Figure \ref{fig:saddle}). Note that the case of an apparent linear law, $\check{\alpha}=1$, is not possible for finite values of $\beta$ and $\alpha=1$, but requires $\alpha=-1$. 

When optimizing the resistances for an entire network, one simply takes the sum of \eqref{eq:lagmu} over all branches. Further including either the condition \eqref{eq:optr}, this yields
\begin{equation}
    \mathcal{M}^*=\sum_i  (\beta k)^{\frac{1}{\beta+1}}(1+(\beta k)^\beta) J_i^{\frac{\beta (\alpha+1)}{\beta+1}},
 \end{equation}
to be coupled to the KCL and the conditions of input and output currents. Note that for $\alpha=1$, the problem is again one of extremizing dissipation, as in Eq. \eqref{eq:redlagpower}, but now for the apparent resistance law with exponent $\check{\alpha}$, 
\begin{equation}
    \mathcal{M}^* =\sum_i \frac{1+\beta}{\beta}\mathcal{D}_i \propto \sum_i J_i^{\check{\alpha}+1},  \ \ \ \ \ \ \ \ \ \  (\alpha=1).
    \label{eq:Mres}
\end{equation}
Differently from the extremization in Eq. \eqref{eq:redlagpower}, here in Eq. \eqref{eq:Mres}, the parameters of the apparent resistance laws \eqref{eq:optr} are all equal and based on the cost. As a result, the currents achieve a maximum uniformity, either in configurations with loops (i.e., all branches active for $\check{\alpha}>1$) or in tree configurations ($\check{\alpha}<1)$, where branches that would give rise to loops are pruned. 

\begin{figure}[!b]
\centering
\includegraphics[width=\linewidth]{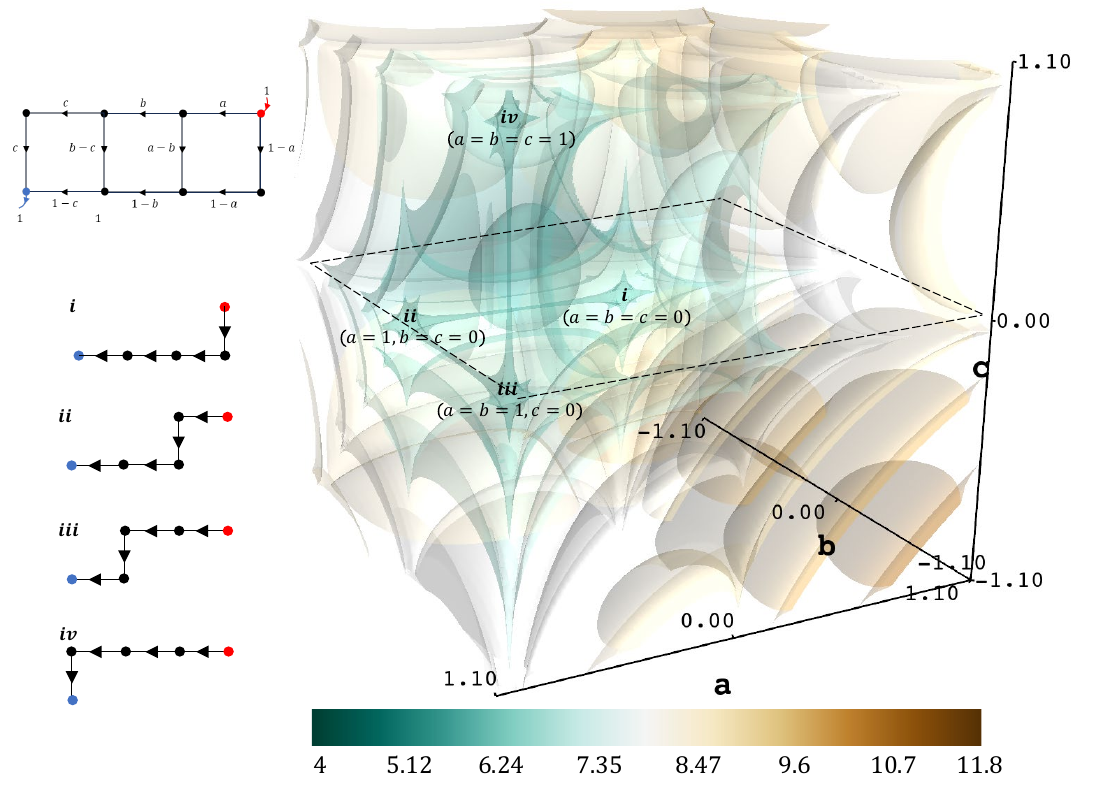}
\caption{\label{fig:optimal_loopsbis}Selection of optimal network configurations for the triple loop case, with unit input in the top left inlet and single outlet at the bottom right. The dissipation is shown for $\gamma=0.5$. $i$, $ii$, $iii$, and $iv$ are the global optimal network configurations.}
\end{figure}

This type of optimization has been used in optimal channel networks \cite{rodriguez1997fractal,banavar1999size,banavar2001scaling,rinaldo2014evolution}, while the corresponding functional, when expressed as a function of resistances, has also been used as a potential for the evolution of transport properties of the type
\begin{equation}
    \frac{d R_i}{dt}=-\frac{\partial \mathcal{M}^*}{\partial R_i}, 
\end{equation}
with artificial time $t$ \cite{burger17mesoscopic}. Similar equations for conductances were previously proposed, using different approaches, by \cite{tero2007mathematical,hu2013adaptation}. Depending on the value of $\beta$, the resulting Lagrangians range from a smooth, single minimum surfaces for the case of congested flow, resulting in networks with loops ($\check{\alpha}<1$) to surfaces with multiple local minima for the case of branched transport, resulting in tree networks (corresponding to the unstable transport law with $\check{\alpha}>1$). In the branched case, the evolutions towards the best configurations (i.e., trees) results in pruning of branches which are out-competed as redundant loops are eliminated to obtain a transport network with given nodes and current inputs/outputs with minimum dissipation, in a manner similar to a discrete Monge-Kantorovich optimal transportation problem \cite{santambrogio2015optimal}.

The functional $\mathcal{M}^*$, coupled to KCL, was analyzed in detail in \cite{banavar2001scaling} and plotted for a single and double loop network. For $\check{\alpha}>1$, the functional is convex with a single minimum corresponding to the most balanced network with loops, while for $\check{\alpha}<1$ it develops multiple singular minima in correspondence of tree networks. The case of the branched transport with $\check{\alpha}=1/2$ is shown in Figure \ref{fig:optimal_loopsbis} for the triple loop case. The complexity of the three-dimensional landscape and the various local minima corresponding to the different tree configurations are evident. The number of possible trees grows fast with the size of the network (e.g., in relation to Kirchhoff's theorem), with many of them being quasi-optimal with self-similar, fractal properties as the size of the domain is increased \cite{rodriguez1997fractal, banavar2001scaling, rinaldo2014evolution}.

\section{Conclusions}

The thermodynamic formalism of nonlinear resistive transport networks proposed here is based on the GTPs, first proposed by Millar in 1951 \cite{millar1951cxvi} and their Legendre-transform symmetry \cite{verhas2014gyarmati}. We stressed the generality of these Lagrangians for complex resistive networks with generic nonlinear transport laws. Such functionals, obtained from virtual variations of currents or forces around the operating point of the network, provide gradient flows that lead to quasi-static evolution criteria toward NESS. 

We discussed the somewhat {\it ad hoc} nature of optimality principles based on dissipation and entropy production, which originate from the proportionality between content and co-content and generalized dissipation in case of power-law resistances. A careful analysis of the role of different constraints helped clarify the type of extremization at the network operating point and the fundamental equivalence of the previously proposed optimization principles (e.g., least dissipation or maximum entropy production). We also showed how such potentials determine the stability of cases with multiple solutions \cite{kondepudi2014modern}, resulting in non-equilibrium phase transitions \cite{ge2009thermodynamic}, described by equations of state similar to those of equilibrium thermodynamics. 

The optimization of transport properties (i.e., resistances), pioneered by Murray \cite{murray1926physiological} and extended by many others \cite{rodriguez1997fractal, banavar2001scaling, bohn2007structure, burger17mesoscopic}, was also linked to a minimization of dissipation for the case of power-law costs. In such cases, the presence of tree or loop structures in the network is directly related to the convexity or concavity of the cost functional \cite{giaccone2025coexistence} and is connected to the problem of optimal transport \cite{santambrogio2015optimal}.

Future work will be devoted to try to formulate an ensemble theory for networks with thermodynamic fluctuations around the operating points (e.g., \cite{porporato2024thermodynamic}, as well as to extend the proposed thermodynamic formalisms to the set of self-similar trees (i.e., the OCNs \cite{rodriguez1997fractal}) and explore the possible similarities with the sub-optimal glassy state \cite{bonetti2020channelization, anand2023self} away from the ground-state optimum \cite{maritan1996universality}.

\bibliographystyle{elsarticle-num}

\bibliography{reference}

\begin{thebibliography}{10}
\bibitem{bandle2017dido}
C.~Bandle, Dido’s problem and its impact on modern mathematics, Notices of the AMS 64~(9).
\end{thebibliography}

\end{document}